\documentclass{aastex}                 
\usepackage{emulateapj5,apjfonts,epsf}
 
\received{date1}
\accepted{date2}
\journalid{}{}
\articleid{}{}

\journalinfo{{\sc The Astrophysical Journal}, 2002, in press}

\slugcomment{Accepted for publication in The Astrophysical Journal
Letters, June 13, 2002}

\shortauthors{Camilo et al.}
\shorttitle{PSR~J1930+1852 in SNR~G54.1+0.3}

\begin{document}

%
%

\def\psr{PSR~J1930+1852}
\def\snr{G54.1+0.3}
\def\chandra{{\em Chandra\/}}
\def\asca{{\em ASCA\/}}

\title{Discovery of a 136 millisecond radio and X-ray pulsar in SNR
G54.1+0.3 }

\author{F.~Camilo,\altaffilmark{1} 
  D.~R.~Lorimer,\altaffilmark{2}
  N.~D.~R.~Bhat,\altaffilmark{3}
  E.~V.~Gotthelf,\altaffilmark{1}
  J.~P.~Halpern,\altaffilmark{1}
  Q.~D.~Wang,\altaffilmark{4}
  F.~J.~Lu,\altaffilmark{4,5} and
  N.~Mirabal\altaffilmark{1} }
\altaffiltext{1}{Columbia Astrophysics Laboratory, Columbia University,
  550 West 120th Street, New York, NY~10027}
\altaffiltext{2}{University of Manchester, Jodrell Bank Observatory,
  Macclesfield, Cheshire, SK11~9DL, UK}
\altaffiltext{3}{NAIC, Arecibo Observatory, HC03 Box 53995, PR~00612}
\altaffiltext{4}{Astronomy Department, University of Massachusetts,
  Amherst, MA~01003}
\altaffiltext{5}{Laboratory of Cosmic Ray and High Energy Astrophysics,
  Institute of High Energy Physics, CAS, Beijing 100039, China}

\begin{abstract}
We report the discovery of a pulsar with period $P = 136$\,ms and
dispersion measure 308\,cm$^{-3}$\,pc in a deep observation of the
supernova remnant (SNR) \snr\ with the Arecibo radio telescope.  Timing
measurements of the new pulsar, J1930+1852, reveal a characteristic age
of $P/2\dot P = 2900$\,yr and spin-down luminosity $\dot E = 1.2 \times
10^{37}$\,erg\,s$^{-1}$.  We have subsequently searched archival
\asca\ X-ray data of this SNR, and detect pulsations with a consistent
period.  These findings ensure that \psr\ is the pulsar that powers the
``Crab-like'' SNR~\snr.  Together with existing \chandra\ observations
of the SNR, we derive an X-ray pulsed fraction (2--10\,keV) of $\approx
27\%$.  We also find that the cooling efficiency of the pulsar wind
nebula (PWN) is intermediate between those of the Vela and Crab PWNe:
$L_{X} \mbox{(2--10\,keV)} \sim 0.002 \dot E$.  \psr\ is a very weak
radio source, with period-averaged flux density at 1180\,MHz of
$60\,\mu$Jy.  For a distance of 5\,kpc, its luminosity, $\sim
1$\,mJy\,kpc$^2$, is among the lowest for known young pulsars.

\end{abstract}


\section{Introduction}\label{sec:intro} 

The supernova remnant (SNR) \snr\ is a close analogue of the Crab
Nebula in several respects.  At radio wavelengths it has a
filled-center morphology, a flat spectrum, and shows significant
polarization \cite{rfa+85,vb88}.  Its X-ray spectrum is non-thermal,
and the X-ray and radio extents are comparable ($\la 2'$; Lu,
Aschenbach, \& Song 2001)\nocite{las01}.  It is a classic compact
synchrotron nebula powered by its putative energetic pulsar, with no
evidence for a thermal component corresponding to shocked ISM or
stellar ejecta.  Using the \chandra\ X-ray Observatory, Lu et
al.~(2002)\nocite{lwa+02} have recently identified the central compact
source (pulsar candidate) and imaged with $\sim$ arcsecond resolution
beautiful coherent structures (e.g., a ring) that are a manifestation
of the relativistic pulsar wind interacting with the ambient medium.

The morphological and spectral properties of \snr\ revealed by
\chandra\ leave no room for doubt as to the presence of a central
pulsar.  It is nevertheless of considerable interest to detect actual
pulsations, and to measure the period of rotation $P$ and its
derivative $\dot P$.  Because the SNR is powered entirely by the
pulsar, knowledge of the input energy source (pulsar spin-down)
luminosity ($\dot E = 4 \pi^2 I \dot P/P^3$, where $I \equiv
10^{45}$\,g\,cm$^2$ is the neutron star moment of inertia) would
improve our understanding of the energetics of the nebula.  Also, the
pulsar characteristic age $\tau_c = P/2 \dot P$ is a useful (if
sometimes crude) measure of the SNR age, especially in cases like
\snr\ where no independent reliable age estimate exists.  Finding radio
and/or X-ray pulsations is also important because a significant sample
of well-studied SNRs and their pulsars provides a window into the
distribution of initial spin periods and magnetic fields of neutron
stars.  Finally, detection of pulsations addresses the poorly
constrained luminosity distribution, ``beaming fraction'', and hence
Galactic population of young neutron stars.

The SNR~\snr\ was searched for a radio pulsar by Gorham et
al.~(1996)\nocite{gra+96} with the ``pre-upgrade'' Arecibo telescope.
At a frequency of 1408\,MHz, their 30\,min observation of a
40\,MHz-wide band reached a claimed sensitivity of 300\,$\mu$Jy for a
pulsar with period $P \sim 0.1$\,s of duty-cycle 10\% and dispersion
measure $\mbox{DM} \sim 300$\,cm$^{-3}$\,pc (based upon their
assumptions, we find their limiting sensitivity to be $\sim
100\,\mu$Jy).  Following the striking \chandra\ observations of \snr\
\cite{lwa+02}, we attempted the most sensitive search currently
possible at Arecibo.  In this Letter we report the discovery of a
136\,ms radio pulsar which, through a subsequent detection in archival
\asca\ X-ray data, is confirmed to be the neutron star powering
SNR~\snr.

\section{Radio search}\label{sec:radio}

On 2001 August 29 we collected data for a full source transit of
2.7\,hr\footnote{Because the source declination is very close to the
observatory latitude, there is a ``hole'' at $| \mbox{zenith angle} | <
1\fdg06$ within which it cannot be tracked; nevertheless we continued
data (noise) collection during this 8\,min interval in order to
maintain phase coherence of the data set.} at a center frequency of
1175\,MHz.  However these data were too corrupted by radio-frequency
interference to be searched, and the observations were redone on 2002
April 29.

The incoming signals from the telescope were recorded using the
Wide-band Arecibo Pulsar Processor (WAPP), a fast-dump digital
correlator (Dowd, Sisk, \& Hagen 2000)\nocite{dsh00} currently capable
of sampling a band of either 50 or 100\,MHz and determining either 3-
or 9-level correlation functions.  All observations for this search
utilized the 100\,MHz 3-level mode of the WAPP in which raw
auto-correlation functions (ACFs) for each of the two polarizations
(IFs) were written to disk as either 32- or 16-bit numbers.  Data from
2001 August were collected in 32-bit mode recording 256 lags every
570\,$\mu$s with the two IFs summed on-line.  Data from 2002 April were
collected in 16-bit mode recording 256 lags every 295\,$\mu$s
separately for each of the two IFs at a frequency of 1180\,MHz.

Off-line preparation of WAPP data for search processing proceeded by
applying the van~Vleck correction (see, e.g., Hagen \& Farley
1973)\nocite{hf73} to each set of 256 ACFs to remove unwanted 3-level
quantization effects.  The resulting unbiased ACFs were then Fourier
transformed to produce spectral channels so that the data following
this step are, in effect, equivalent to power streams from a filter
bank with $2\times 256$ spectral channels.  Since the search targeted a
young ($P\ga30$\,ms) pulsar, the ideal sampling interval for our
observations would have been $\sim 1$\,ms.  However the slowest sample
time of the WAPP (determined by the size of the registers used to store
accumulated correlation values) is $\approx 400\,\mu$s in 16-bit, 2-IF
mode, and the faster-sampled data recorded were instead decimated
off-line by adding every 8 time samples before further processing.  At
this point summing of the 2 IFs and reduction in precision to 8-bit
quantities were also performed.  The data reduction tools used for this
stage of the analysis are described by Lorimer
(2001)\nocite{lor01b}\footnote{Available at
http://www.jb.man.ac.uk/$\sim$drl/sigproc.}.

The data from 2002 April were reduced thereafter in standard fashion.
The 256 time series of $2^{22}$ samples, each decimated to a resolution
of 2.36\,ms, were de-dispersed at 1001 trial DMs in the range
0--1000\,cm$^{-3}$\,pc.  Finally, these 1001 time series were each
searched for periodic signals over a range of duty cycles with an
FFT-based code (see Lorimer et al.~2000 for details)\nocite{lkm+00}.

An unmistakable periodic and dispersed signal with $P \approx
136.8$\,ms was detected with maximum signal-to-noise ratio of 20.7 at
$\mbox{DM} \approx 308$\,cm$^{-3}$\,pc.  Nine days later we confirmed
this signal in a shorter observation, by which time the barycentric
period had increased by $\sim 0.6\,\mu$s, indicating a period
derivative $\dot P \sim 8\times 10^{-13}$.  At this stage, we searched
archival X-ray data for (and found) pulsations (\S~\ref{sec:xray}).  We
have begun timing observations of the pulsar and so far have obtained
times-of-arrival (TOAs) on 5 occasions spanning 24 days.  In addition
we reduced, and obtained TOAs from, the 2001 August data.  We have used
the {\sc tempo}\footnote{See http://pulsar.princeton.edu/tempo.} timing
software and the TOAs, along with the pulsar position known to $\sim
0\farcs6$ accuracy from \chandra\ data \cite{lwa+02}, to derive the
$P$ and $\dot P$ listed in Table~\ref{tab:parms}.  We
regard this as a preliminary ephemeris because of the 8 month gap
between the first TOA and the recent set.  

\medskip
\epsfxsize=8.0truecm
\epsfbox{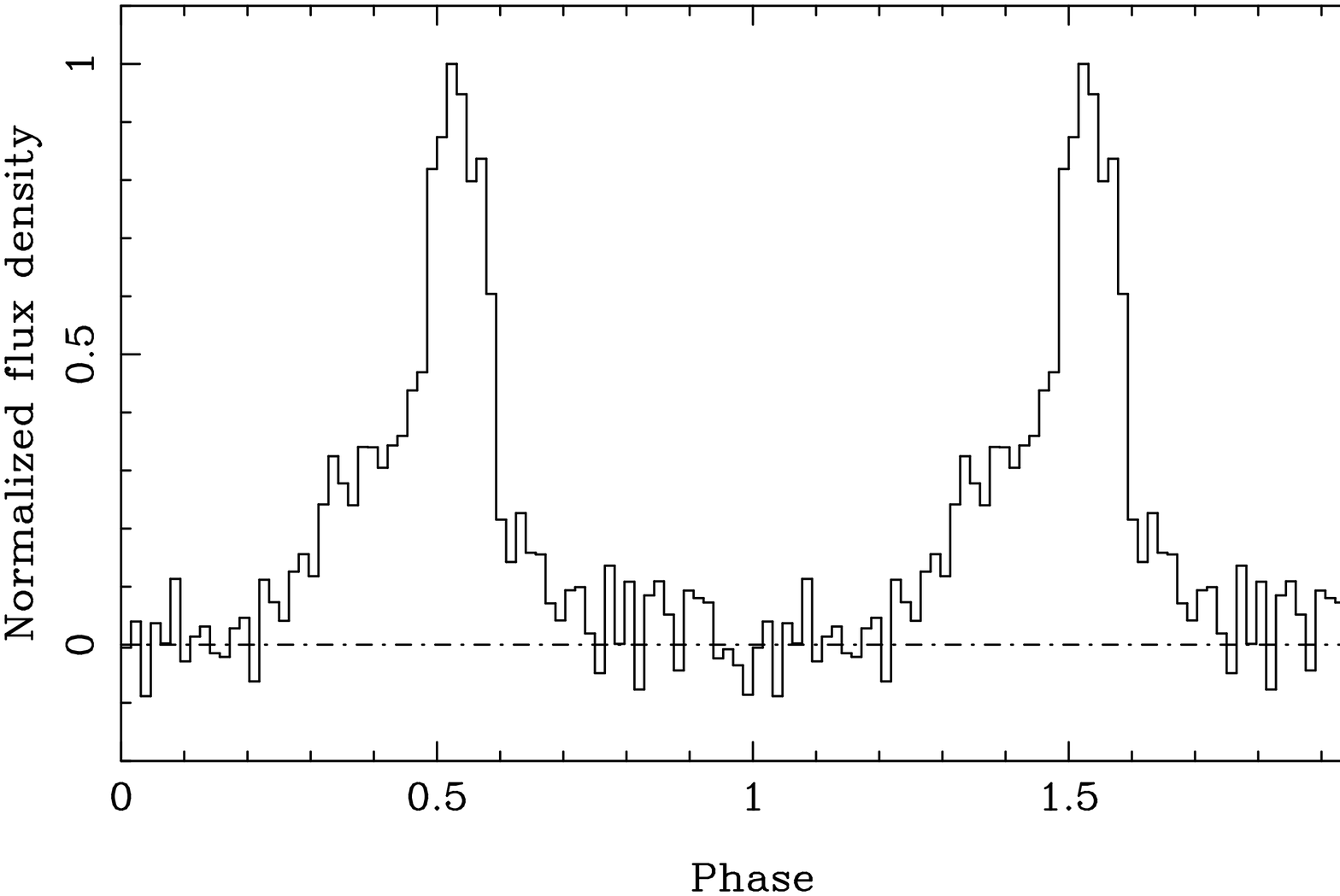}
\figcaption[f1.ps]{\label{fig:radioprof} 
Average radio pulse profile of \psr\ at 1180\,MHz, based on 8.75\,hr of
Arecibo data.  Notice the broad ``wings'' of emission.  Phase zero is
arbitrary. }

\medskip

The average pulse profile at a frequency of 1180\,MHz is displayed in
Figure~\ref{fig:radioprof}, and was obtained by phase-aligning and
summing the profiles obtained on each day according to the ephemeris
given in Table~\ref{tab:parms}.  We have measured the area under each
of the individual profiles, and their baseline offsets, and used these,
together with the known telescope gain and system temperature
(including their dependencies on zenith angle and the contribution from
Galactic synchrotron and SNR emission), to estimate the average flux
density of the pulsar.  We obtain $S_{1180} = 60 \pm 10\,\mu$Jy, where
the uncertainty is the addition in quadrature of the standard deviation
of the 6 measurements ($6\,\mu$Jy) and a $\sim 10$--15\% contribution
accounting for imperfectly known system parameters.

\section{X-ray pulsar}\label{sec:xray}

The field containing \snr\ was observed in the X-ray band on 1997 April
27--28 with the {\em Advanced Satellite for Cosmology and
Astrophysics\/} (\asca; Tanaka, Inoue, \& Holt 1994)\nocite{tih94} to
search for a Crab-like pulsar.  This observation was designed to allow
temporal studies with the two gas scintillation imaging spectrometers
(GISs) at a time resolution of 0.488\,ms throughout the exposure.  To
ensure consistent (high) temporal resolution, during periods when
real-time telemetry was not available, data were taken at a reduced
($1/4$) spatial resolution at the slower (medium) bit-rate and stored
on board.  Although the spatial resolution of the GIS ($\sim 2'$) is
insufficient to resolve the pulsar from the surrounding nebula, the
moderate ($\approx 5$--10\%) energy resolution in the 0.5--10\,keV
bandpass allows some measure of energy discrimination of the
instrumental and X-ray background.  To search for pulsations, we
combined GIS data from both instruments and from data sets taken in
different spatial modes.  The latter necessitated rebinning the sky
pixels of the high bit-rate data to match the medium bit-rate data
set.  All data were edited to exclude times of high background
contamination using the standard (REV2) screening criteria.  The
resulting effective observation time was 22.7\,ks spanning 43.1\,ks.
Photon arrival times were corrected to the solar system barycenter
using the JPL DE200 ephemeris and the known pulsar position
(Table~\ref{tab:parms}).


Arrival times were obtained for 2210 photons extracted from a $3'$
radius aperture centered on the peak of X-ray emission from \snr\ and
restricted in energy to 2--10\,keV to optimize the signal-to-noise
ratio.  We generated a periodogram using the $\chi^2$ statistic by
folding the extracted photons over a range of test frequencies centered
on the extrapolated radio measurement, using a crude spin-down rate
based on the discovery and confirmation observations.  We detected a
single $\approx 6\,\sigma$ signal with a narrow pulse profile near the
predicted period.  Using the $Z_n^2$ test to gain sensitivity to $n$
harmonics of the fundamental frequency \cite{bbb+83}, we obtained a
best signal with $Z_3^2 = 47.8$.  While not significant in a blind
search (probability of chance occurrence is $1.3 \times 10^{-8}$ per
trial, or 0.13 in a 0--20\,Hz search range with $2\,\mu$Hz resolution),
this is the strongest signal in the search range.  Impressively, the
barycentric period of $P = 136.74374(5)$\,ms on 50566.0 MJD differs by
only ($0.15\pm0.05$)\,$\mu$s from the period extrapolated back to this
epoch with the preliminary radio timing ephemeris obtained 5\,yr later
(\S~\ref{sec:radio}).  This confirms that the X-ray signal represents a
bona fide detection of pulsations from the radio pulsar.  As there is
no X-ray source in the \chandra\ field other than the point-source
pulsar candidate that could generate the pulsed X-ray emission detected
by \asca, we conclude unambiguously that we have in fact detected radio
and X-ray pulsations from \psr\ at the center of SNR~\snr.


The X-ray pulse profile, shown in Figure~\ref{fig:xrayprof}, is similar
in shape to the radio profile (Fig.~\ref{fig:radioprof}), with a single
narrow peak of $\sim 20\%$ duty-cycle.  The pulse width is slightly
larger than that found in the radio, but is limited by the statistics
of the observation.  The measured pulsed fraction, defined as the ratio
of counts in the light-curve above the mean baseline in the off-pulse
interval to the total counts, is 9.5\%.  However this result is
contaminated by emission from the nebula.  Using all available data
(\asca\ GIS and \chandra\ ACIS-S3) together we can constrain further
the properties of the pulsar.

\medskip
\epsfxsize=8.0truecm
\epsfbox{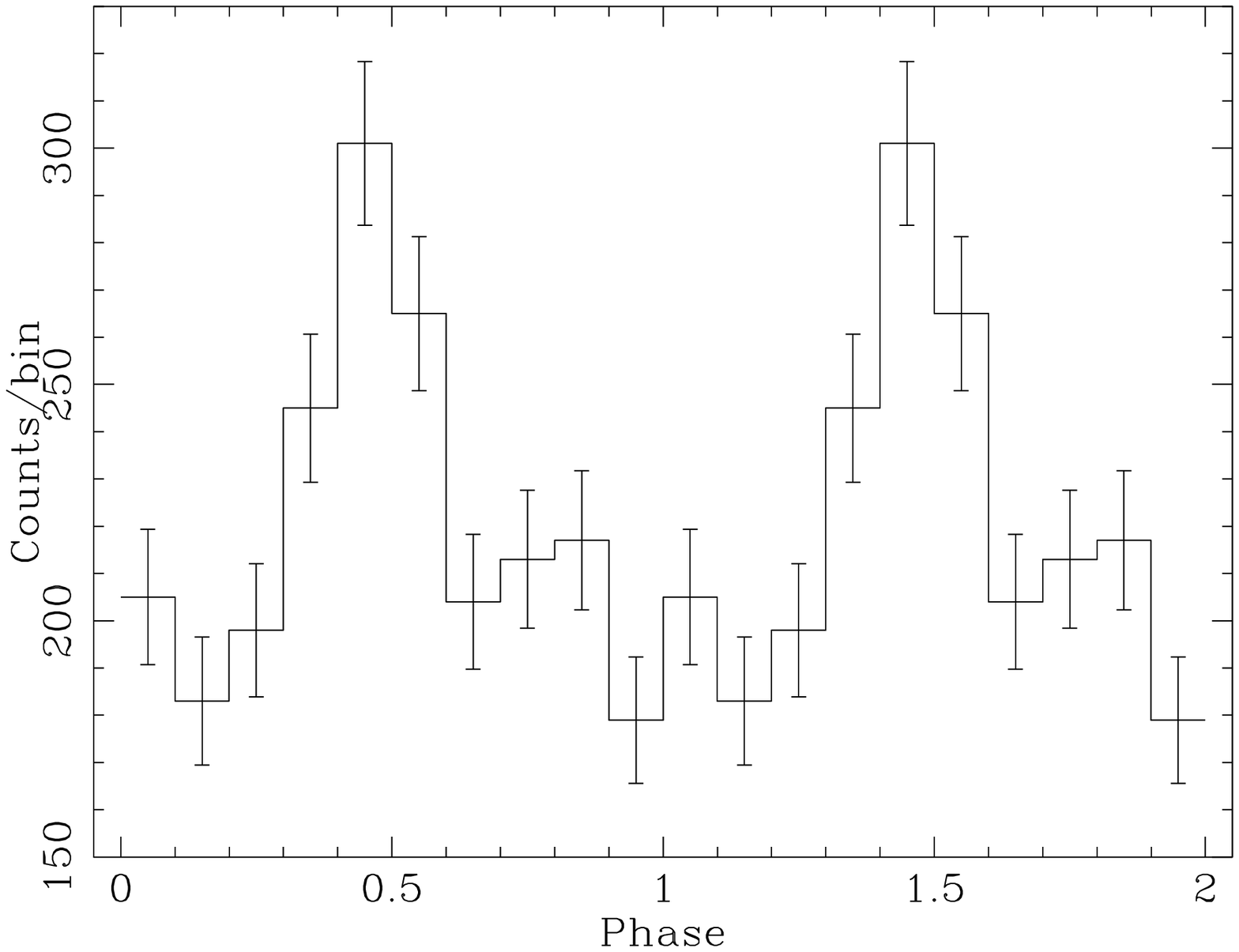}
\figcaption[f2.ps]{\label{fig:xrayprof} 
X-ray pulse profile of \psr\ in the 2--10\,keV band from \asca\ GIS.
Accounting for diffuse emission resolved with a \chandra\ observation
(see text), pulsed fraction is $\approx 27\%$.  Phase zero is
arbitrary. }

\medskip

With a count rate of $0.064 \pm 0.002$\,s$^{-1}$, the point source in
the \chandra\ observation of Lu et al.~(2002)\nocite{lwa+02} was
significantly affected by pile-up \cite{dav01}.  We have used a new
version of the spectral analysis software {\sc xspec} that accounts for
this, obtaining a corrected pulsar photon index $\Gamma =
1.35_{-0.10}^{+0.06}$ (uncertainties are 90\% confidence levels) and
measured flux in the 2--10\,keV band of $1.7 \times
10^{-12}$\,erg\,cm$^{-2}$\,s$^{-1}$.  The absorbing column density was
fixed at $N_H = 1.6 \times 10^{22}$\,cm$^{-2}$ \cite{lwa+02}, for which
the unabsorbed pulsar flux is $1.9 \times
10^{-12}$\,erg\,cm$^{-2}$\,s$^{-1}$.  This spectrum corresponds to a
count rate in the 2--10\,keV \asca\ GIS band of 0.044\,s$^{-1}$,
computed with {\sc wpimms}\footnote{See
http://heasarc.gsfc.nasa.gov/Tools/w3pimms.html.} for a $6'$ radius
aperture.  The pulsed rate measured within this aperture is
0.012\,s$^{-1}$, implying a pulsed fraction of $\approx 27\%$.  This
compares to a fraction of $\approx 7\%$ for the Vela pulsar (Helfand,
Gotthelf, \& Halpern 2001)\nocite{hgh01}, and $\ga 75\%$ for
PSR~J2229+6114 \cite{hcg+01}, two young pulsars with values of $\dot E$
within a factor of 2 of \psr's.  The absolute phase alignment of the
X-ray and radio profiles (to be determined from scheduled {\em RXTE\/}
and planned radio observations) will provide a further diagnostic of
the emission mechanism(s).

Pulsars such as this one are known to emit a substantial fraction of
their spin-down luminosity in magnetospheric $\gamma$-rays.  \psr\ lies
$1\fdg4$ from the high energy ($>100$\,MeV) EGRET $\gamma$-ray source
3EG~J1928+1733 \cite{hbb+99}.  This is slightly outside the 99\%
confidence-level error box, but given its weakness and the possibility
of an extended or multiple source, an association is not out of the
question.

\section{Discussion}\label{sec:disc}

The $P$ and $\dot P$ measured for \psr\ imply a large spin-down
luminosity $\dot E = 1.2 \times 10^{37}$\,erg\,s$^{-1}$, small
characteristic age $\tau_c = 2900$\,yr, and surface magnetic dipole
field strength $B = 3.2\times10^{19} (P \dot P)^{1/2} = 1.0 \times
10^{13}$\,G (Table~\ref{tab:parms}).  These parameters are virtually
identical to those of PSR~J1124$-$5916 in SNR~G292.0+1.8, and place it
in the group of $\sim 10$ pulsars with the highest values of $\dot E$
and smallest apparent ages known, all of which are associated with SNRs
(see Camilo et al.~2002a)\nocite{cmg+02}.  However its parameters are
substantially different from those of the Crab pulsar, which remains
the only Galactic neutron star known with $P<50$\,ms and $\dot E >
10^{38}$\,erg\,s$^{-1}$.

The pulsar's characteristic age allows for a useful estimate of its
actual age (and that of the SNR), given by $\tau = 2 \tau_c [ 1 -
(P_0/P)^{n-1} ]/(n-1)$ under the assumption of constant magnetic
moment, where $P_0$ is the initial period and $n$ is the braking index
of rotation \cite{mt77}.  For magnetic dipole braking, $n = 3$, and
measured values span $2 \la n \la 3$ (see, e.g., Camilo et
al.~2000)\nocite{ckl+00}.  Initial periods of rotation are thought to
range from $\approx 10$\,ms, to perhaps $\ga 90$\,ms (for
PSR~J1124$-$5916; Camilo et al.~2002a)\nocite{cmg+02}.  Considering the
extremes of $n \approx 2$ and $P_0 \approx 90$\,ms in turn, the age of
J1930+1852/\snr\ likely lies in the range 1500--6000\,yr.

The distance of SNR~\snr\ has been estimated from a measurement of the
X-ray absorption column density, which is found to be about half the
total Galactic absorption in this direction.  As the Galaxy here
extends to $\sim 10$\,kpc from the Sun, Lu et al.~(2002)\nocite{lwa+02}
consider $d \sim 5$\,kpc.  Assuming the SNR to be associated with the
star-forming region G53.9+0.3, Velusamy \& Becker (1988)\nocite{vb88}
obtain $d \sim 3.2$\,kpc.  The dispersion measure of
\psr\ (Table~\ref{tab:parms}) provides an independent estimate: the
Taylor \& Cordes (1993)\nocite{tc93} free electron density/distance
model suggests $d \sim 12$\,kpc.  A newer model incorporating, among
other improvements, discrete regions of enhanced ionized material
(J.~M. Cordes \& T.~J. Lazio 2002, in preparation), yields $d \la
8$\,kpc (T.~J. Lazio 2002, private communication), consistent with the
estimate derived from X-ray measurements.  While the distance to this
SNR/pulsar pair remains uncertain, we here retain $d \sim 5$\,kpc as a
plausible estimate.

We now comment on some features of the pulsar wind nebula (PWN).  A
composite \chandra\ image of \snr\ (Fig.~\ref{fig:cxo}) demonstrates
the trend of X-ray spectral softening from the inner to the outer
regions of the nebula (as first indicated in the analysis of Lu et
al.~2002)\nocite{lwa+02}.  This is likely caused by a combination of
synchrotron cooling and adiabatic expansion of the shocked wind
material.  The 2--10\,keV luminosity of the PWN/pulsar combination is
$L_X = 2.2 \times 10^{34} (d/5\,\mbox{kpc})^2$\,erg\,s$^{-1} \sim 0.002
\dot E$.  This is a factor of $\sim 2$ lower than that of the PWN
surrounding PSR~J1124$-$5916, a pulsar with identical spin parameters
to \psr\ \cite{cmg+02}\footnote{Likewise the radio luminosity
($10^7$--$10^{11}$\,Hz) of \snr, $L_R \sim 1 \times 10^{33}
(d/5\,\mbox{kpc})^2$\,erg\,s$^{-1} \sim 0.0001 \dot E$ \cite{vb88}, is
half that of the PWN powered by PSR~J1124$-$5916.  However, in one
significant respect these two pulsar/SNR systems are different:
PSR~J1124$-$5916 and its PWN are part of the text-book composite
SNR~G292.0+1.8, embedded in a large and bright shell of stellar ejecta
\cite{hsb+01,prh+02}, whereas to date no thermal emission has been
detected surrounding \snr.}.  By contrast, the cooling efficiencies
($L_X/\dot E$) of the PWNe surrounding the Vela \cite{pksg01} and
J2229+6114 \cite{hcg+01} pulsars, with similar $\dot E$, are much
lower: $L_X \la 10^{-4} \dot E$; while that of the Crab Nebula is $\sim
25$ times higher (e.g., Helfand \& Becker 1987)\nocite{hb87}.  The
extent to which these differences are due to different ambient media,
energy spectra of the injected pulsar wind (including the time
evolution thereof), and ages, among other variables, is not clear.

\medskip
\epsfxsize=8.0truecm
\epsfbox{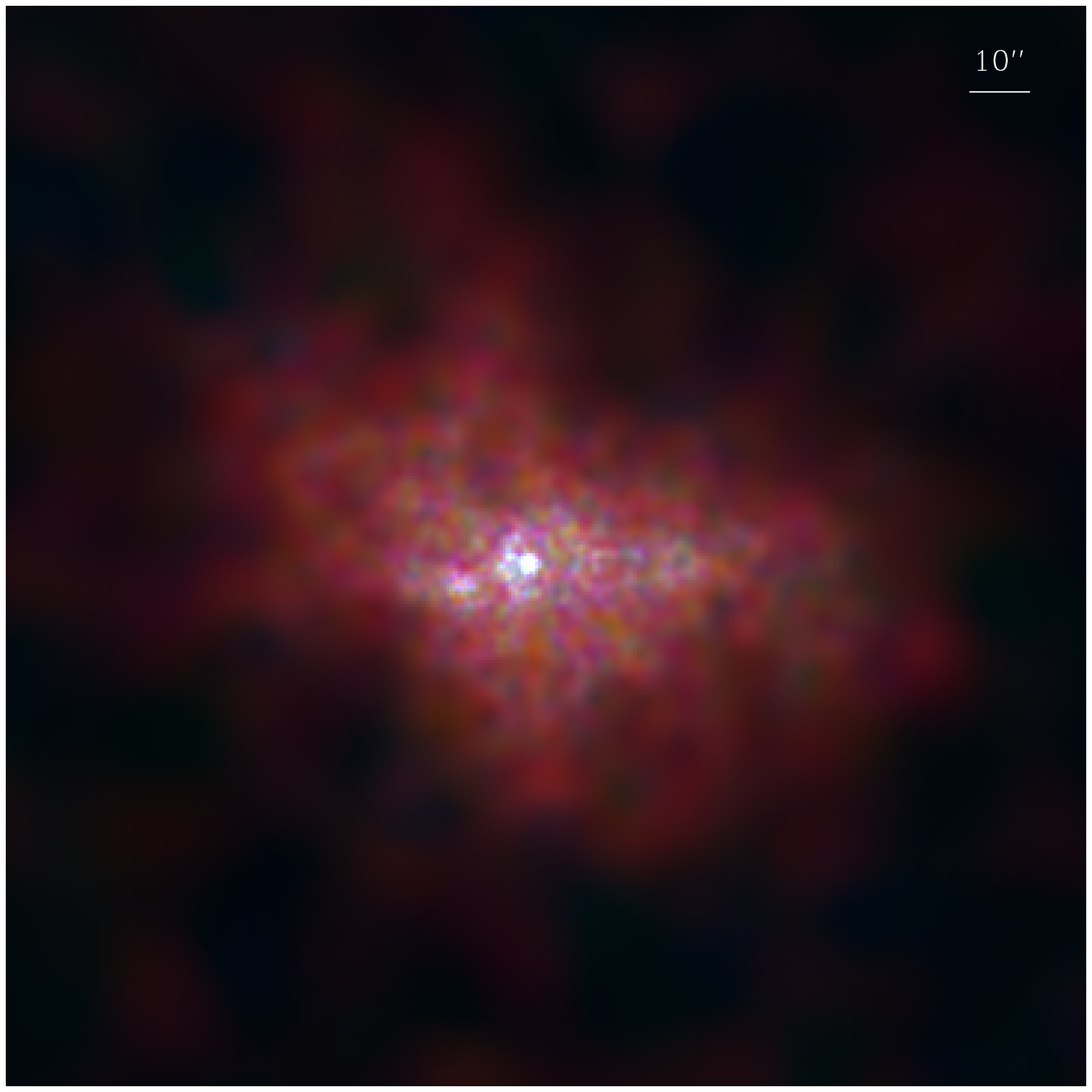}
\figcaption[f3.ps]{\label{fig:cxo} 
\chandra\ ACIS-S3 image of SNR~\snr, color-coded by energy:
1.0--2.0\,keV (red), 2.0--3.5\,keV (green), and 3.5--8.0\,keV (blue).
The X-ray images in individual bands were adaptively smoothed in
identical manner with a Gaussian filter in order to achieve a
count-to-noise ratio of $\approx 6$ in the final 1--8\,keV image.
North is to the top and East is to the left. }

\medskip

With the discovery of \psr, four very young and energetic pulsars have
been uncovered in the past year, all with luminosities of $\sim
1$\,mJy\,kpc$^2$ in the 1400\,MHz radio band (Halpern et al.~2001;
Camilo et al.~2002a, b; this Letter)\nocite{hcg+01,cmg+02,csl+02}.
These are well below the luminosities of previously known young pulsars
(see Camilo et al.~2002a)\nocite{cmg+02}, and it is now clear that such
pulsars can be extremely faint.  Determining their intrinsic luminosity
distribution requires disentangling observed ``pseudo-luminosities''
from beam-averaged values.  For example, do the broad wings of emission
in \psr\ visible in Figure~\ref{fig:radioprof} indicate nearly aligned
rotation and magnetic axes with an impact angle grazing the outer
boundary of a radio beam with possibly large averaged luminosity?
Studies of polarized emission may provide important clues to constrain
the beaming geometry of individual pulsars.

Naturally, a large sample of young pulsars is also desirable, together
with upper limits from the most sensitive searches possible of
well-selected targets.  In this regard, it is significant that new and
upgraded radio telescopes are now available.  The discovery of \psr\ is
an excellent example, as it could only barely have been made with the
Arecibo telescope prior to its upgrade in the late 1990s: in addition
to the increased bandwidth, we can now sample a lower-frequency band
(1175 vs. 1400\,MHz), where pulsars are brighter and the telescope's
gain is higher.  Planned improvements in system temperature and
bandwidth at Arecibo, the new Green Bank Telescope, and the remarkably
productive Parkes telescope, along with the continued availability of
the superb \chandra\ X-ray Observatory, offer the prospect of yet
further progress in studies of faint young neutron stars.

\acknowledgments

We thank Joe Taylor and Joel Weisberg for generously giving us some of
their telescope time for the original radio observation, and John
Harmon for scheduling it.  We are also grateful to Shri Kulkarni for
designing the \asca\ observation, and to Jeff Hagen, Bill Sisk and Andy
Dowd for their sterling work in realizing the potential for wide-band
pulsar observations with the upgraded Arecibo telescope.  We
acknowledge useful discussions on beaming geometry with Michael
Kramer.  The Arecibo Observatory is part of the National Astronomy and
Ionosphere Center, which is operated by Cornell University under a
cooperative agreement with the National Science Foundation.  This work
was funded in part by grants from SAO (GO1-2063X: FC; GO1-2068X: QDW \&
FJL) and NASA (LTSA NAG~5-7935: EVG).  DRL is a University Research
Fellow funded by the Royal Society.  FJL is partially supported by the
Special Funds for Major State Basic Research Projects of China.


%
%
\begin{deluxetable}{ll}
\tablecaption{\label{tab:parms}Parameters of \psr\ }
\tablecolumns{2}
\tablewidth{0pc}
\tablehead{
\colhead{Parameter}   &
\colhead{Value}     \\}
\startdata
R.A. (J2000)\dotfill                           & $19^{\rm h}30^{\rm m}30\fs13$\\
Decl. (J2000)\dotfill                          & $+18\arcdeg52'14\farcs1$     \\
Period, $P$ (ms)\dotfill                       & 136.855046957(9)             \\
Period derivative, $\dot P$\dotfill            & $7.5057(1)\times10^{-13}$    \\
Epoch (MJD [TDB])\dotfill                      & 52280.0                      \\
Dispersion measure, DM (cm$^{-3}$\,pc)\dotfill & $308(4)$                     \\
Flux density at 1180\,MHz ($\mu$Jy)\dotfill    & $60 \pm 10$                  \\
Pulse FWHM at 1180\,MHz (ms)\dotfill           & $15 \pm 2$                   \\
Pulse FWHM at 2--10\,keV (ms)\dotfill          & $\sim 25$                    \\
Distance of SNR~\snr, $d$ (kpc)\dotfill        & $\sim 5$                     \\
Derived parameters:                                      &                    \\
~~Characteristic age, $\tau_c$ (yr)\dotfill              & 2900               \\
~~Spin-down luminosity, $\dot E$ (erg\,s$^{-1}$)\dotfill & $1.2\times10^{37}$ \\
~~Magnetic field strength, $B$ (G)\dotfill               & $1.0\times10^{13}$ \\
~~Luminosity at 1400\,MHz (mJy\,kpc$^2$)\dotfill         & $\sim 1$           \\
\enddata
\end{deluxetable}

\end{document}